\documentclass{jfm}
\usepackage{graphicx}
\usepackage{epstopdf, epsfig}
\usepackage{color}

\shorttitle{DNS of Taylor-Couette flow with vertical asymmetric rough walls}
\shortauthor{F Xu, J Su, B Lan, P Zhao, Y He, C Sun and J Wang}

\title{Direct numerical simulation of Taylor-Couette flow with vertical asymmetric rough walls}

\author{Fan Xu\aff{1,2},
  Jinghong Su\aff{5,6},
  Bin Lan\aff{1,2},
  Peng Zhao\aff{2,3},
  Yurong He\aff{1}
  \corresp{\email{rong@hit.edu.cn}},
  Chao Sun\aff{5,6}\corresp{\email{chaosun@tsinghua.edu.cn}},
 \and Junwu Wang\aff{2,3,4}\corresp{\email{jwwang@ipe.ac.cn}}}

\affiliation{\aff{1}School of Energy Science and Engineering, Harbin Institute of Technology, Harbin, 150001, P. R. China
\aff{2}State Key Laboratory of Multiphase Complex Systems, Institute of Process Engineering, Chinese Academy of Sciences, P. O. Box 353, Beijing 100190, P. R. China
\aff{3}School of Chemical Engineering, University of Chinese Academy of Sciences, Beijing, 100049, P. R. China
\aff{4}Innovation Academy for Green Manufacture, Chinese Academy of Sciences, Beijing 100190, P. R. China
\aff{5}Center for Combustion Energy, Key Laboratory for Thermal Science and Power Engineering of Ministry
of Education, Department of Energy and Power Engineering, Tsinghua University, Beijing 100084, P. R. China
\aff{6}Department of Engineering Mechanics, School of Aerospace Engineering, Tsinghua University, Beijing 100084, P. R. China}

\begin{document}

\maketitle

\begin{abstract}
Direct numerical simulations are performed to explore the effects of the rotating direction of the vertically asymmetric rough wall on the transport properties of Taylor-Couette (TC) flow, up to a Taylor number of \textit{Ta} = 2.39$\times$10$^{7}$. It is shown that, compared to the smooth wall, the rough wall with vertical asymmetric strips can enhance the dimensionless torque \textit{Nu}$ _{\omega}$. More importantly, at high \textit{Ta}, clockwise rotation of the inner rough wall (where the fluid is sheared by the steeper slope side of the strips) results in a significantly greater torque enhancement compared to counter-clockwise rotation (where the fluid is sheared by the smaller slope side of the strips), due to the larger convective contribution to the angular velocity flux. However, the rotating direction has a negligible effect on the torque at low \textit{Ta}. The larger torque enhancement caused by the clockwise rotation of the vertically asymmetric rough wall at high \textit{Ta} is then explained by the stronger coupling between the rough wall and the bulk, attributed to the larger biased azimuthal velocity toward the rough wall at the mid-gap of the TC system, the increased turbulence intensity manifested by larger Reynolds stress and a thinner boundary layer, and the more significant contribution of the pressure force on the surface of the rough wall to the torque.

\end{abstract}


\section{Introduction}

Turbulent flows with rough walls are ubiquitous in nature, and many engineering applications must contend with rough boundaries. The viscous length scales in the flow decrease with increasing Reynolds numbers, and eventually, every surface appears to be rough, even when the roughness is small in absolute scale. Nikuradse (1933) was the first to study how local wall roughness (sand glued to the wall) affects global transport properties in pipe flow. Since then, numerous studies \citep{chan2018secondary,rouhi2019direct,ma2020scaling,modesti2021dispersive,jelly2022impact} and reviews \citep{flack2014roughness,chung2021predicting} have explored the effects of wall roughness in (pipe or channel) turbulence. Instead of utilizing open-channel or pipe flow with rough walls, we employ a Taylor-Couette (TC) apparatus, which is a closed system with an exact balance between energy input and dissipation. Furthermore, due to its simple geometry and excellent controllability, the TC system offers favorable conditions for both numerical and experimental investigations \citep{zhu2018wall,verschoof2018rough}.

In most experimental and numerical studies, both of the inner and outer cylinders of TC system are smooth surfaces (see \cite*{grossmann2016high} for a comprehensive review). The effects of rough walls have only been investigated in recent decades. According to the shape of rough walls, they can be divided into three categories. The first type of rough wall is the irregular rough surface made by adhering particles randomly on the cylindrical wall \citep{berghout2019direct,berghout2021characterizing,bakhuis2020controlling}, it was found that the torque can be enhanced by the irregular rough wall, indicating drag enhancement. The second type of rough wall is that the regular roughness is arranged in the way aligned with the mean flow, which is called \textquoteleft{parallel roughness}\textquoteright. It was found that the parallel grooves result in drag enhancement at relatively high Taylor numbers once the height of roughness is larger than the velocity boundary layer (BL) thickness \citep{zhu2016direct}, this is because the plumes are ejected from the tips of these grooves and the system forms a secondary circulating flow inside the groove. Stronger plume ejections have an enhanced effect on the torque and then lead to drag enhancement. On the other hand, the parallel corrugated surface resulted in drag reduction at low Taylor number \textit{Ta}, whereas drag enhancement was found at high \textit{Ta} \citep{ng2018interaction,razzak2020numerical}. Similar findings were also reported in studies employing micro-grooves \citep{razzak2020numerical,xu2023effect}.

The third type is to arrange the roughness perpendicular to the mean flow, i.e., \textquoteleft{vertical roughness}\textquoteright. \cite{cadot1997energy} first reported this rough wall effect on drag by attaching vertical ribs on the inner and outer cylinders. Inspired by their work, \cite{van2003smooth} performed further experiments with the same style of roughness by conducting four groups of experiments, i,e., two smooth walls, rough-inner/smooth-outer, smooth-inner/rough-outer, and two rough walls. Both studies found that the vertical roughness has a drag enhancement effect on the TC flow due to the extra torque of the rough elements coming from pressure force. \cite{zhu2017disentangling} carried out a quantitative analysis on the origins of torque at the rough wall and found that the contribution of pressure force to the torque at the rough wall is of prime importance for drag enhancement. Other works \citep{lee2009experimental,motozawa2013experimental,zhu2018torque,verschoof2018rough,sodjavi2018effects} also studied the effects of vertical rough walls, focusing on the effects of the number of vertical strips, the strip height, and/or the radius ratio. Up to date, all studies on the effects of regular rough walls have used symmetrical, rough walls, resulting in the same influences by different rotating directions of the cylinders,
whereas the asymmetric effect of different rotating directions with vertical asymmetrical rough walls on the TC flow, which may make a huge difference, remains unexplored.

In this article, direct numerical simulations (DNS) of TC flow with vertical asymmetrical rough walls were carried out to study how different rotating directions affect the global response as well as local flow behavior.
The manuscript is organized as follows. In \S 2, the numerical method and settings are described. In \S 3, the relationships between Nusselt number and Taylor number for vertical asymmetrical rough inner walls with different heights and rotating directions are presented, and the mechanism behind the differences in torque is explained. The local flow behavior is also analyzed. Finally, conclusions are drawn in \S 4.

\section{Numerical method and setting}\label{sec2}
\begin{figure}
	\centering{\includegraphics[width=1\textwidth]{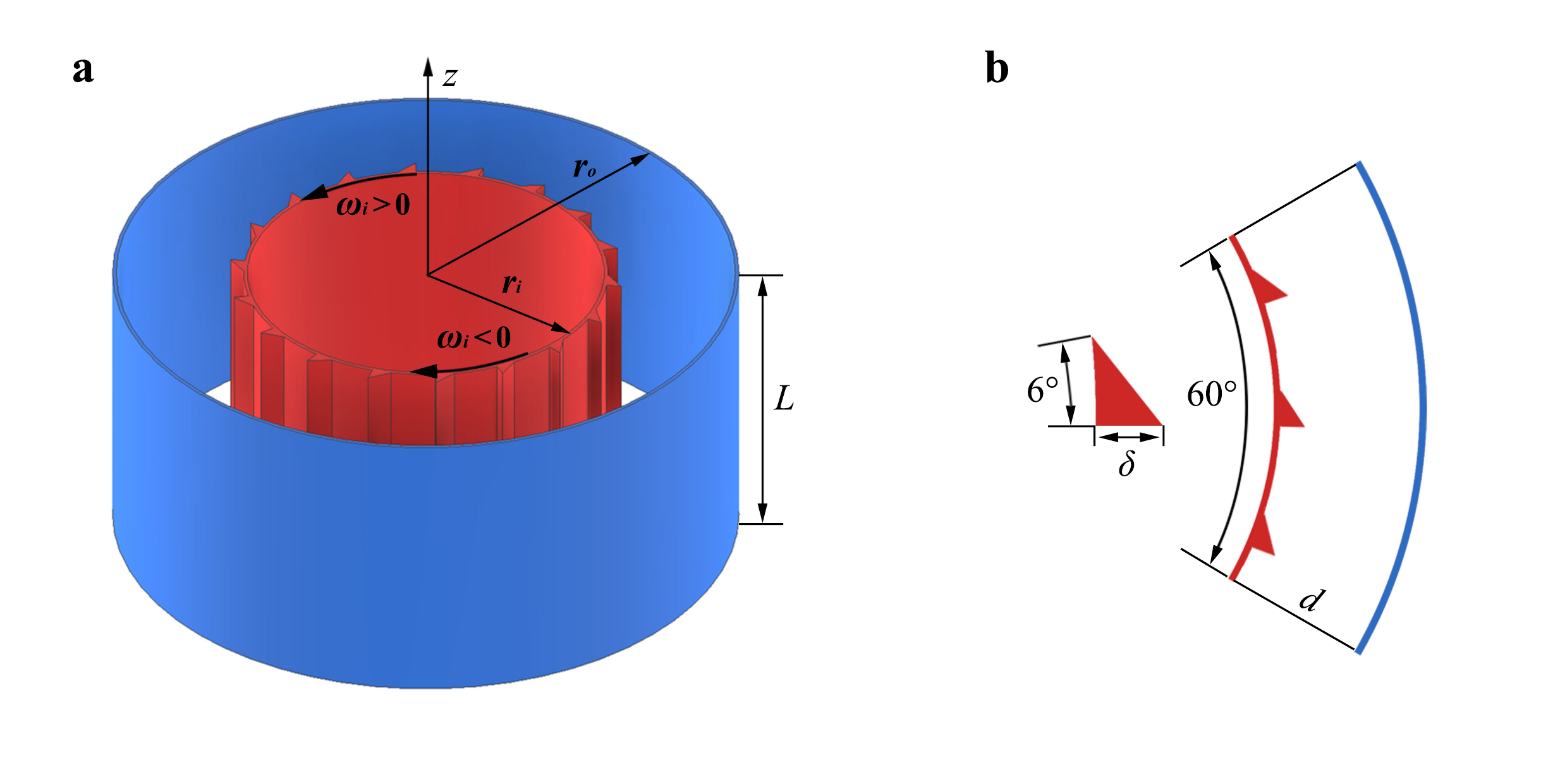}}
	\caption{Schematic view of the Taylor-Couette system and the geometry of roughness. (\textit{a}) Three-dimensional view. The inner cylinder with radius \textit{r}$_{i}$ is rotating with angular velocity $ \omega_{i} $. $\omega$$ _{i} $ $ > $ 0 represents that the fluid is sheared by the smaller slope side of the strips, referred subsequently as counter-clockwise rotation. $\omega$$ _{i} $ $ < $ 0 indicates the fluid is sheared by the steeper slope side of the strips, referred subsequently as clockwise rotation. The outer cylinder with radius \textit{r}$_{o}$ is stationary. (\textit{b}) Cross-section view of the gap between the two cylinders: \textit{d} = \textit{r}$ _{o} $$ - $\textit{r}$ _{i} $. The rough elements are eighteen triangular vertical strips positioned equidistantly on the inner cylinder wall. The height of the rough elements are 0.1\textit{d} and 0.2\textit{d}. In present simulations, a rotational symmetry of six is used. Therefore, the computational domain contains 1/6 of the azimuthal width and has three rough elements on the inner cylinder.
		}
	\label{TCsystem}
\end{figure}

In the present study, the outer cylinder is at rest, the inner cylinder is rotating and thus driving the flow. The outer cylinder is a smooth wall, and the inner one is a rough wall, with both walls subject to the no-slip boundary condition. Axially periodic boundary conditions are used, meaning that present study does not include the effects of end walls presented in the TC experiments. The inner cylinder is roughened by attaching eighteen vertical strips of right triangle cross section where one side of the strips is perpendicular to the inner wall and the strips with a height of $ \delta $ are equally distributed in the azimuthal direction (see figure \ref{TCsystem}). The motivation behind this study is to investigate the impact of asymmetric roughness on torque in the Taylor-Couette system, particularly its effects on turbulent statistics. While there are various possibilities for asymmetric roughness shapes, we focus on the right triangular rib (the simplest asymmetric geometry model) as a starting point to examine the effects on statistics of turbulent Taylor-Couette flow. Our primary objective is to investigate how the presence of asymmetric roughness elements impacts both global transport and local flow statistics in Taylor-Couette turbulence.

For the sake of simplicity and ease of explanation, as illustrated in Figure \ref{TCsystem}(a), the angular velocity of inner cylinder $\omega$$ _{i} $ $ > $ 0 represents that the fluid is sheared by the smaller slope side of the strips, referred subsequently as \textquoteleft{counter-clockwise rotation}\textquoteright. In contrast, $\omega$$ _{i} $ $ < $ 0 indicates the fluid is sheared by the steeper slope side of the strips, referred subsequently as \textquoteleft{clockwise rotation}\textquoteright. 
The gap width (\textit{d}) is calculated as the difference between the radii of the outer cylinder (\textit{r}$ _{o} $) and the inner cylinder (\textit{r}$ _{i} $). The radius ratio is $\eta$ = \textit{r}$_{i}$/\textit{r}$_{o}$ = 0.714 and the aspect ratio is $ \varGamma $ = \textit{L}/\textit{d} = 2$ \pi $/3, where \textit{L} is the length of axial periodicity. The geometry of the system is fixed at the radius ratio of \textit{$\eta$} = 0.714 and the outer cylinder is stationary, to make a direct comparison with previous results \citep{ostilla2013optimal,xu2022direct}. With $ \varGamma $ = 2$ \pi $/3, we can have a relatively small computational domain with a pair of Taylor vortices. A rotational symmetry of six is selected to reduce the computational cost while not affecting the results, which has been verified by previous studies \citep{brauckmann2013direct,ostilla2015effects,xu2022direct}. As a result, there are only three triangular strips in the azimuthal direction, as shown in figure \ref{TCsystem}(b).

The fluid between the two cylinders is assumed to be Newtonian and incompressible. The motion of the fluid under these assumptions is governed by the continuity equation
\begin{equation}\label{6}
	\nabla\cdotp\textbf{\textit{u}}=0,
\end{equation}
and the momentum conservation equation \citep{zhu2016direct},
\begin{equation}
	\frac{\partial\textit{\textbf{u}}}{\partial\textit{t}}+\nabla\cdot(\textbf{\textit{uu}})=-\nabla\textit{p}+\frac{f(\eta)}{Ta^{1/2}}\nabla^{2}\textbf{\textit{u}},
\end{equation}
where \textbf{\textit{u}} and \textit{p} are the dimensionless fluid velocity and pressure, respectively. The equations are normalized using the gap width \textit{d}, and the tangential velocity of the inner cylinder \textit{u}$ _{i}$ = \textit{r}$_{i} $$ \omega $$_{i}$, time is normalized by the characteristic length and  velocity \textit{d}/\textit{u}$ _{i}$, and the pressure term is normalized by the square of inner wall velocity and density $ \rho $\textit{u}$_{i}^{2}$. We also define the non-dimensional radius \textit{r}$ ^{\ast} $ to be \textit{r}$ ^{\ast} $ = (\textit{r}$ - $\textit{r}$ _{i} $)/\textit{d}.
\textit{f}($ \eta $) is a geometrical factor written in the form \citep{ostilla2013optimal,zhu2016direct},
\begin{equation}
	f(\eta)=\frac{(1+\eta)^{3}}{8\eta^{2}}.
\end{equation}
The Taylor number can characterize the driving TC flow, in the case of static outer cylinder, it is defined as \citep{grossmann2016high},
\begin{equation}\label{1}
	Ta = {\frac{(1+\eta)^{4}}{64\eta^{2}}}{\frac{d^{2}(r_{i}+r_{o})^{2}\omega_{i}^{2}}{\nu^{2}}},
\end{equation}
where $ \nu $ is the kinematic viscosity of the fluid. An alternative way to determine the system is using the inner Reynolds number that is defined as \textit{Re}$ _{i} $ = \textit{r}$ _{i} $${\omega}$$ _{i} $\textit{d}/$ \nu $, and these two definitions can be interconverted using the formula $ Ta = [f(\eta)Re_{i}]^{2} $. Both the Reynolds and Taylor numbers are presented in table 1. Moreover, the use of the Taylor number, instead of the Reynolds number, is common for distinguishing different TC flow regimes \citep{ostilla2013optimal,ostilla2014exploring,grossmann2016high}. Besides, we also provide the Reynolds number for the roughness based on the average azimuthal velocity at the height of the roughness, which is defined as $ Re_{\delta} = \delta\overline{u}_{\varphi, r = r_{i}+\delta}/\nu $ and shown in table 1 in the appendix.

In TC flow, the angular velocity flux from the inner cylinder to the outer cylinder is strictly conserved along the radius \textit{r}  \citep{eckhardt2007torque}, it is defined as
\begin{equation}\label{2}
	J^{\omega}=r^{3}(\langle\textit{u}_{r}\omega\rangle _{A,t}-\nu\partial_{r}\langle\omega\rangle_{A,t}),
\end{equation}
where \textit{u}$_{r}$ is the radial velocity, $\omega$ is the angular velocity, and $ \langle...\rangle $$_{A,t}$ denotes averaging over a cylindrical surface (averaging over the axial and azimuthal directions) with constant distance from the axis and over time. Here, the radius is selected to be within the scope of \textit{r$ _{i} $}+$ \delta $ $ \le $ \textit{r} $ \le $ \textit{r$ _{o} $}. \textit{J}$ ^{\omega} $ is connected to the dimensionless torque Nusselt number $Nu_{\omega}$ via
\begin{equation}\label{3}
	Nu_{\omega}=J^{\omega}/J^{\omega}_{lam},
\end{equation}
where \textit{Nu}$ _{\omega} $ is the key response parameter in TC flow and \textit{J}$ ^{\omega}_{lam} $ = 2$ \nu $$ r_{i}^{2} $$ r_{o}^{2} $$ \omega_{i} $/($ r_{o}^{2} $$ - $$ r_{i}^{2} $) is the angular velocity flux of the nonvortical laminar state.
Note that \textit{Nu}$ _{\omega} $ can be connected to the experimentally measurable torque $ \tau $ via $ \tau $ = 2$ \pi $\textit{l}$ \rho $\textit{Nu}$ _{\omega} $\textit{J}$ ^{\omega}_{lam} $ by keeping the cylinder rotating with a constant velocity \citep{grossmann2016high}, where \textit{l} is the height of the part of the cylinder on which the torque is measured, $ \rho $ is the fluid density.

Equations (2.1) and (2.2) are solved using a second-order-accuracy, colocated finite-volume method in the Cartesian coordinate system, using OpenFOAM as the computational platform. During the simulations, the results in the Cartesian coordinate are transformed to the format in the cylindrical coordinate, and the simulations are run for at least 40 large eddy turnover times (\textit{d}/\textit{r}$ _{i} $\textit{$ \omega $}$ _{i} $) for data analysis.
The no-slip boundary condition of the inner rough wall was dealt with a second-order-accuracy immersed boundary method \citep{zhao2020cfd,zhao2020Acomputational}. The temporal term is discretized using the second-order backward scheme and the convective term is discretized using a second-order total variation diminishing (Vanleer) scheme. All simulations are achieved using a fixed time step based on the Courant-Friedrichs-Lewy (CFL) criterion and The CFL number is less than 1.0 in all simulations. More details of the simulation accuracy are shown in the appendix.

\begin{figure}
	\centering
	{\includegraphics[width=1\textwidth]{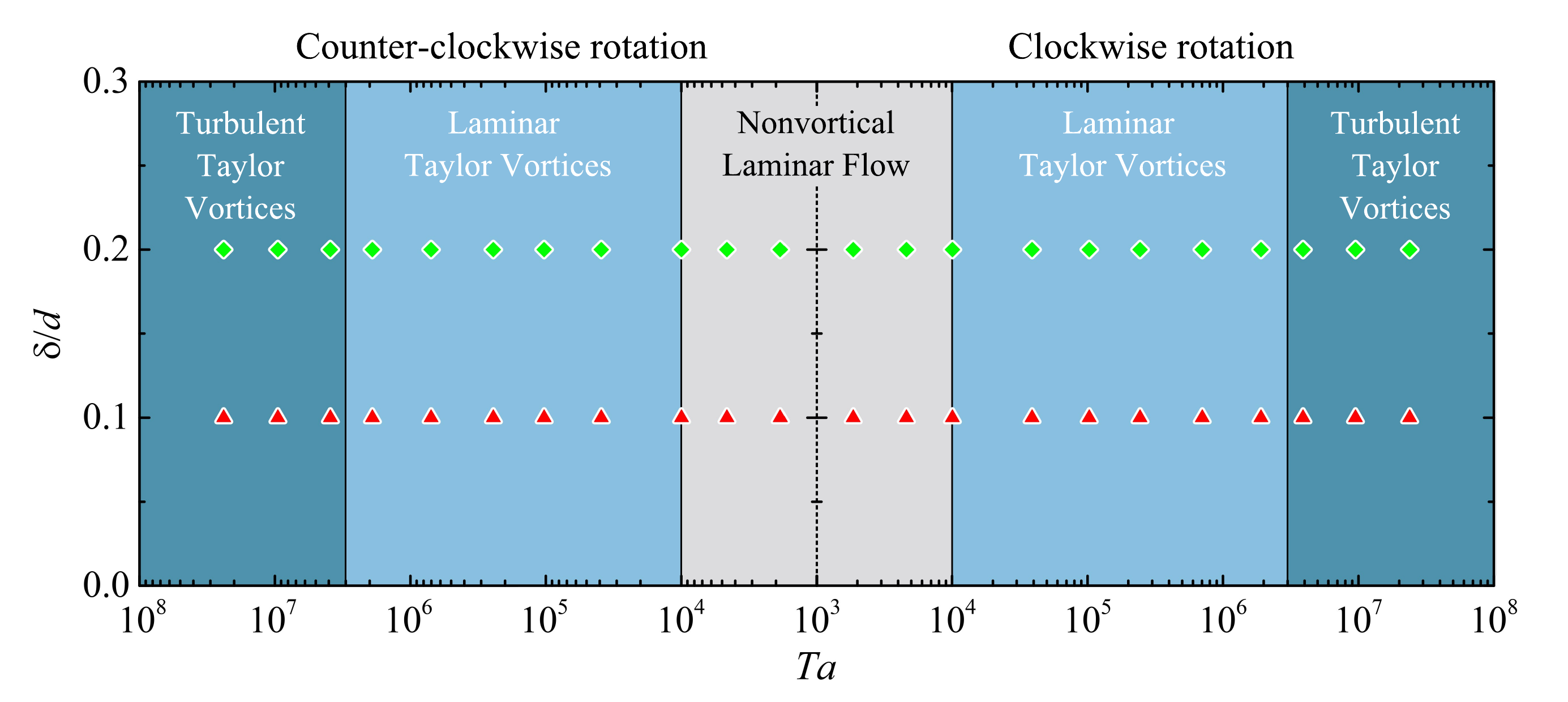}}
	\caption{Explored (\textit{Ta}, $ \delta $/\textit{d}) parameter space. There are three different regimes in the phase spaces, i.e., the nonvortical laminar flow, the laminar Taylor vortices and the turbulent Taylor vortices regimes. The triangles represent the strip height of 0.1\textit{d}, the rhombuses show the strip height of 0.2\textit{d}.}\label{explordspace}
\end{figure}

Two different strip heights ($ \delta $ = 0.1\textit{d} and $ \delta $ = 0.2\textit{d}) on the inner cylinder with different rotating directions, i.e., $\omega$$ _{i} $ $ > $ 0 (counter-clockwise rotation) and $\omega$$ _{i}$ $ < $ 0 (clockwise rotation), were analyzed. In each series with the same strip height, \textit{Ta} ranges from 10$ ^{3} $ to 10$ ^{7} $ or \textit{Re$ _{i} $} is varied from 35 to 3960. The parameter space consists of the Taylor number \textit{Ta} and the strip height $ \delta $/\textit{d} are shown in figure \ref{explordspace}. Note that the vertical solid lines in figure \ref{explordspace} are the transition values of \textit{Ta} for smooth surfaces in TC flow, and the flow states are $ \delta $-dependent. The division of flow state is at smooth surfaces and radius ratio $ \eta $ = 0.714, which is in accordance with the classification methods proposed by \cite{ostilla2014exploring} and \cite{grossmann2016high}, i.e, the determination of the critical Taylor number was based on the onset of Taylor vortices within the TC system.

\section{Results}
\subsection{Dimensionless torque}
\begin{figure}
	\centering
	{\includegraphics[width=1\textwidth]{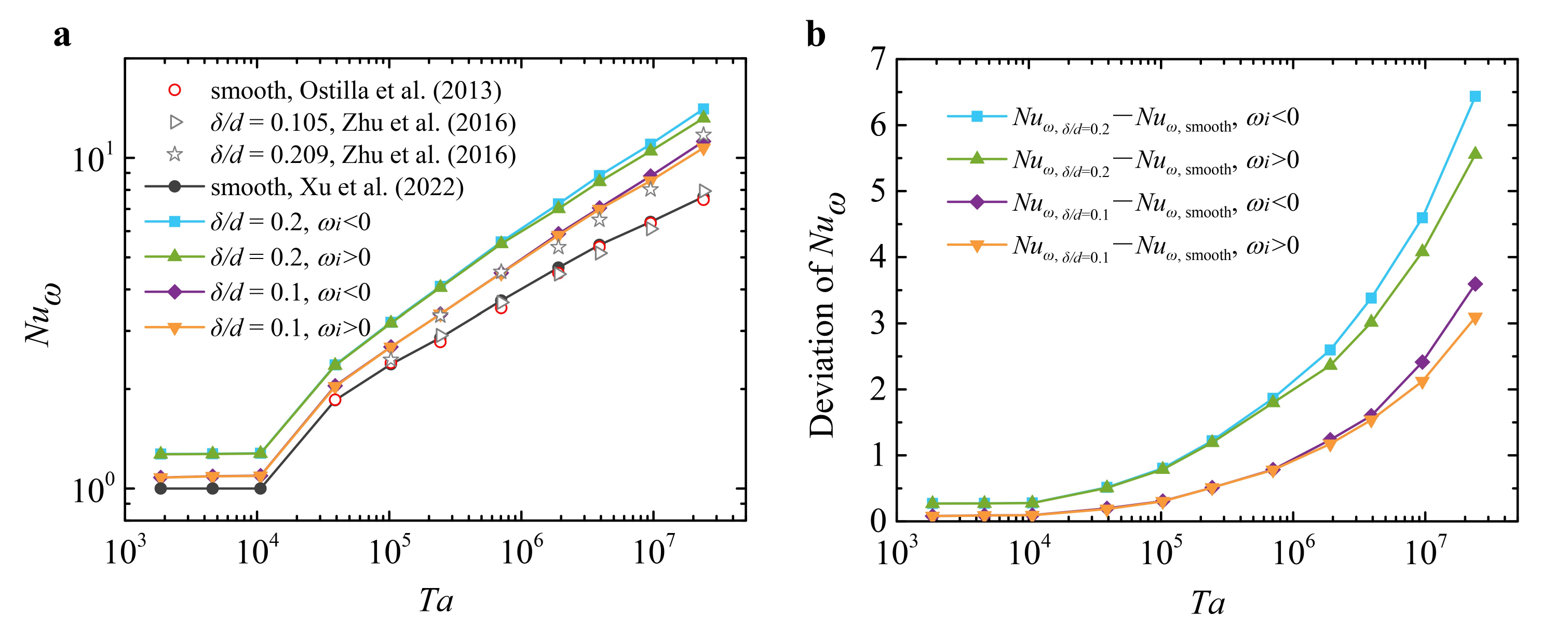}}
	\caption{(a) Nusselt number as a function of \textit{Ta} for smooth cases and rough cases with two strip heights for different rotating directions of the inner cylinder at $ \eta $ = 0.714. The results of previous smooth walls \citep{ostilla2013optimal,xu2022direct} and parallel roughness walls with two different groove heights \citep{zhu2016direct} are also shown for reference. (b) The \textit{Nu}$ _{\omega} $ deviation between rough and smooth walls with increasing \textit{Ta}.}\label{NuvsTa}
\end{figure}

To study the effect of the triangle strip walls, the dimensionless torque \textit{Nu}$ _{\omega} $ is presented as a function of \textit{Ta} (i.e. \textit{Nu}$ _{\omega} $=\textit{ATa}$ ^{\beta} $).
Figure 3(a) shows the dimensionless torque \textit{Nu}$ _{\omega} $ with increasing \textit{Ta} for smooth wall and rough wall with two strip heights rotating in different directions. The results of previous smooth walls \citep{ostilla2013optimal} and parallel roughness walls \citep{zhu2016direct} are also shown in figure \ref{NuvsTa}(a) for reference. In the nonvortical laminar flow regime, the flow only has an azimuthal velocity component and \textit{Nu}$ _{\omega} $ = 1 for smooth wall by definition. But the values of \textit{Nu}$ _{\omega} $ are larger than 1 for rough wall, and the higher the strip, the larger the \textit{Nu}$ _{\omega} $. Although both the flow for smooth and rough cases are purely azimuthal at this regime, the $ \omega $-gradient of the latter is larger and the radial velocity \textit{u}$ _{r} $ = 0. According to equation (2.5), the angular velocity flux \textit{J}$ ^{\omega} $ for the rough walls is larger, i.e., the torque is larger. 
Besides, we also find that the critical Taylor number ($ Ta_{c} $), determined based on the onset of Taylor vortices in the TC system, is affected by the rough surface. We conducted a series of simulations with various strip heights ($ \delta $ = 0.1$ d $ and $ \delta $ = 0.2\textit{d}) at different Taylor numbers. The critical Taylor number ($ Ta_{c} $ $ \approx $ 1.15$ \times $10$ ^{4} $ or $ Ta_{c} $ $ \approx $ 1.35$ \times $10$ ^{4} $ ) was identified as the value at which Taylor vortices became evident. While $ Ta_{c} $ $ \approx $ 1$ \times $10$ ^{4} $ for smooth wall in previous studies \citep{grossmann2016high,xu2022direct} with the same radius ratio $ \eta $ = 0.714, which means that the presence of a rough surface influences the value of this critical Taylor number. These results can be easily understood, the appearance of strips enlarges the effective radius of the inner cylinder, which makes the effective radius ratio of the rough wall larger than that of the smooth wall, therefore, the critical Taylor number is larger \citep{pirro2008direct}.

After the onset of Taylor vortices, no matter whether walls are smooth or not, the torque \textit{Nu}$ _{\omega} $ increases with \textit{Ta}. It is difficult to directly compare our results with other turbulent flow systems with rough walls, but the study with other types of rough walls in TC can be chosen for comparison. As shown in figure \ref{NuvsTa}(a), despite the types of the roughness are discrepant, a similar conclusion is obtained, that is, higher roughness results in a larger torque. But compared with the parallel roughness, the drag enhancement of the vertical one is better. In addition to this, the rotating direction of the rough inner cylinder has no effect on the torque at a fixed strip height for low Taylor numbers. This can be seen more clearly in figure \ref{NuvsTa}(b), which presents the deviation of \textit{Nu}$ _{\omega} $ from the corresponding smooth one at different strip heights and rotating directions. It is shown that the effect of strip height on the torque becomes more significant with increasing \textit{Ta} after the onset of Taylor vortices. And the higher the strip, the larger the torque increase with the same Taylor number. In addition, the effects of rotating direction on the torque for different strip heights are different. For the cases of $ \delta $ = 0.2\textit{d}, the influence of rotating direction on torque appears when \textit{Ta} $ > $ 10$ ^{6} $, and compared with the case of $ {\omega}_{i} $ $ > $ 0, the torque of $ {\omega}_{i} $ $ < $ 0 is larger. But the effect of rotating direction on the torque comes out until \textit{Ta} $\approx$ 10$ ^{7} $ for $ \delta $ = 0.1\textit{d}, the difference between the counter-clockwise and clockwise rotations on the torque of $ \delta $ = 0.1\textit{d} is smaller than the corresponding difference of $ \delta $ = 0.2\textit{d} at same \textit{Ta}.

\begin{figure}
	\centering
	{\includegraphics[width=0.7\textwidth]{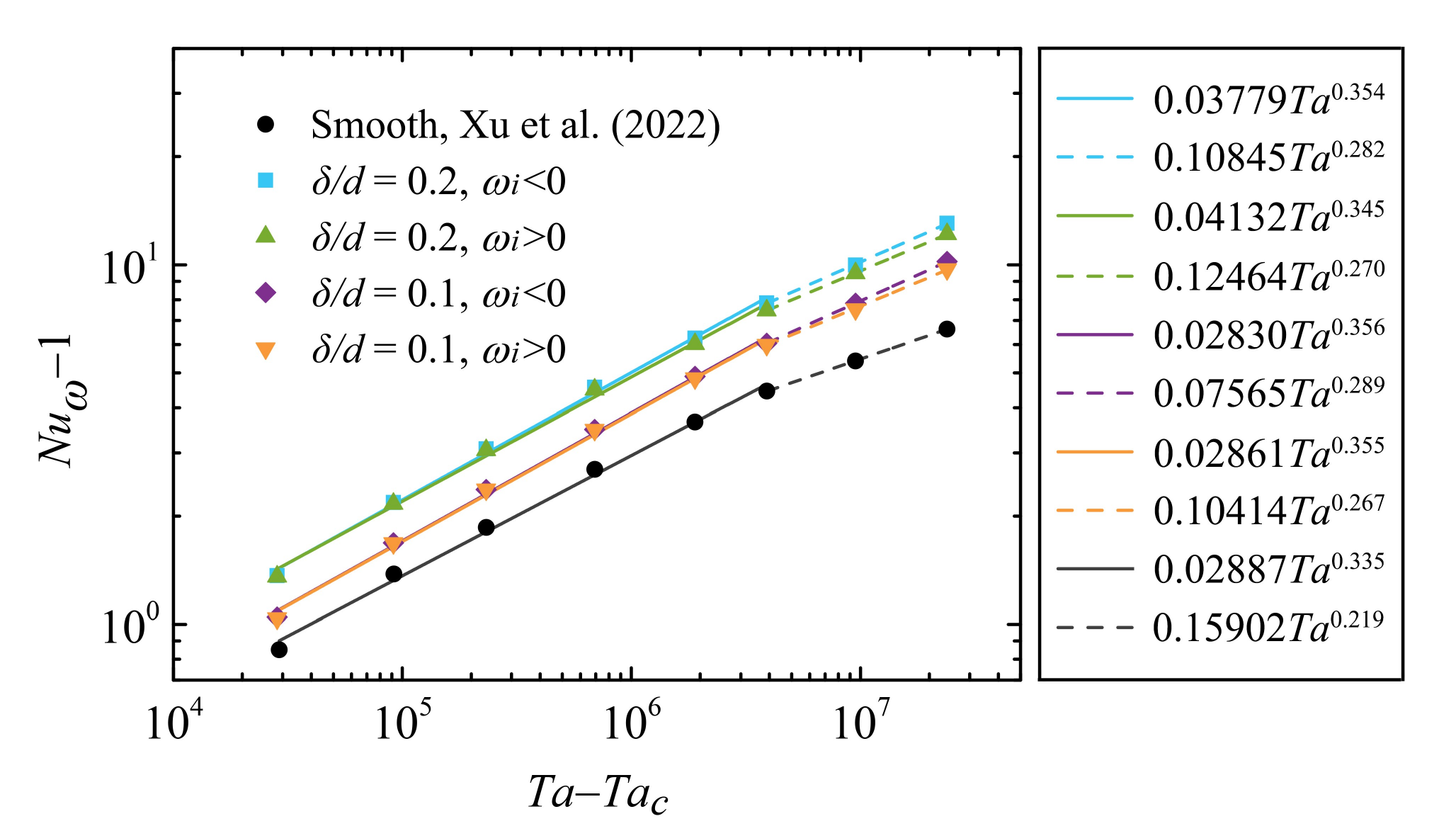}}
	\caption{(a) Nusselt number as a function of \textit{Ta} for smooth cases and rough cases with two strip heights  for different rotating directions of the inner cylinder at $ \eta $=0.714. The data of rough cases are from the present study, and the data of smooth cases are from our previous study \citep{xu2022direct}.}\label{NuvsTa1}
\end{figure}

\textit{Nu}$ _{\omega} $$-$1 is the additional transport of angular velocity on the top of the nonvortical laminar transport in TC flow. Figure \ref{NuvsTa1} shows the numerically calculated \textit{Nu}$ _{\omega}$$-$1 with increasing \textit{Ta} after the appearance of Taylor vortices. Here, we plot \textit{Nu}$ _{\omega} $$ - $1 versus \textit{Ta}$ - $\textit{Ta$ _{c} $} rather than versus \textit{Ta} to better show the scaling at low \textit{Ta} \citep{ostilla2013optimal}. For the smooth TC flow, from \textit{Ta} = 3.9$ \times $10$ ^{4} $ up to \textit{Ta} = 3$ \times $10$ ^{6} $, an effective scaling law of \textit{Nu}$ _{\omega} $$ - $1 $ \sim $ (\textit{Ta}$ - $$\textit{Ta}_{c}$)$ ^{1/3} $ is found, which is connected with the laminar Taylor vortices regime. When \textit{Ta} $ > $ 3$ \times $10$ ^{6} $, there is a transitional region in which the bulk becomes turbulent but the large-scale coherent structure can still be identified when looking at the time-averaged quantities, which is associated with the turbulent Taylor vortices regime \citep{ostilla2014turbulence}. In this transitional regime, the boundary layers are laminar first and become gradually turbulent with increasing \textit{Ta}.

The situation becomes different for the TC flow with rough walls. As shown in figure \ref{NuvsTa1},
at the laminar Taylor vortices regime the effective scaling exponent $ \beta $ $\approx$ 0.35 for different strip heights on the inner cylinder ($ \delta $ = 0.1\textit{d} and $ \delta $ = 0.2\textit{d}) rotating in clockwise ($ {\omega}_{i} $ $ < $ 0) and counter-clockwise ($ {\omega}_{i} $ $ > $ 0) directions. But the situation becomes more complicated at the turbulent regime, the exponent $ \beta $ is influenced not only by the height of strip, but also by the rotating direction of inner rough wall. Figure \ref{NuvsTa1} shows that the effect of strip height on the exponent does not show any regularity, but the influence of rotating direction of the inner rough wall is regular, i,e., the exponent $ \beta $ is slightly larger for the cases of $ {\omega}_{i} $ $ < $ 0, compared to the values of $ {\omega}_{i} $ $ > $ 0 at the same strip height.

\begin{figure}
	\centering
	{\includegraphics[width=1\textwidth]{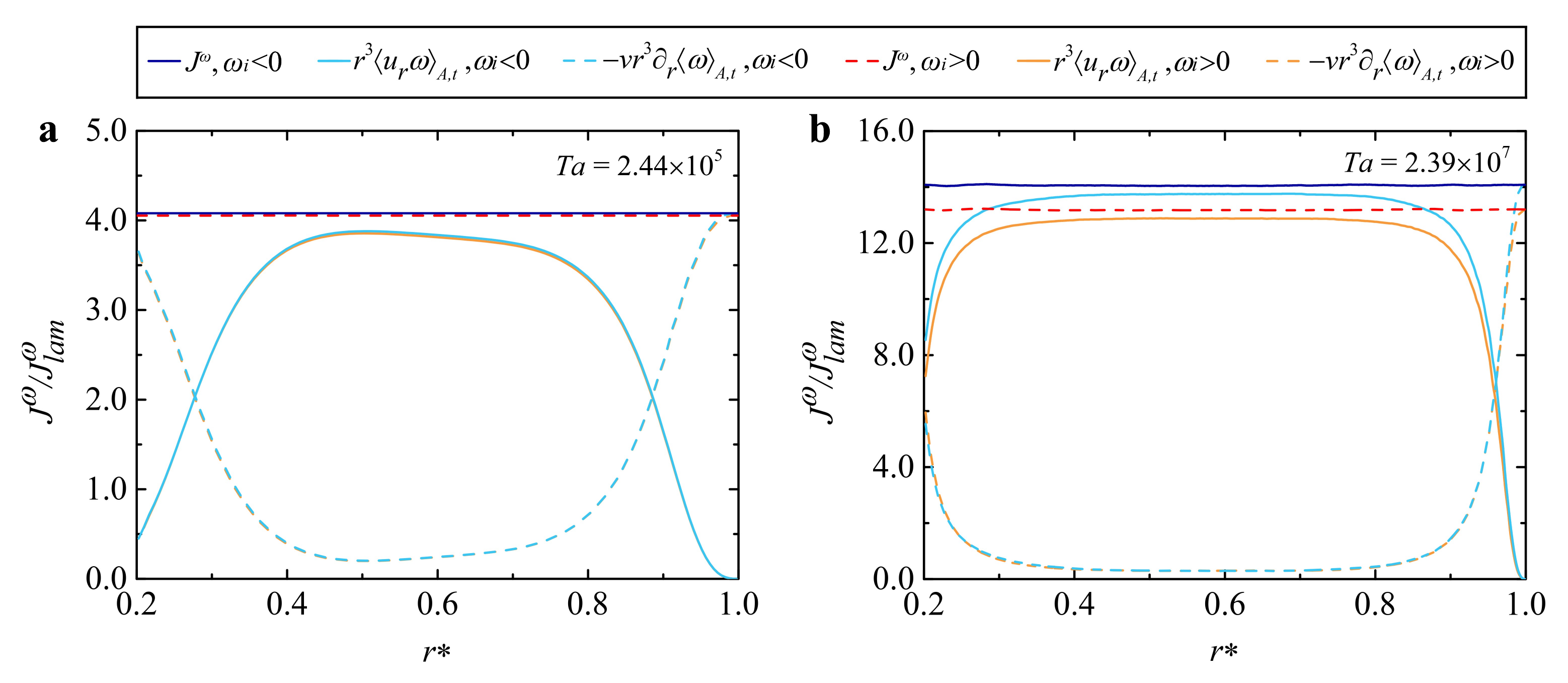}}
	\caption{The convective and diffusive contributions to the angular velocity flux for different rotating directions with $ \delta $ = 0.2\textit{d} at (a) \textit{Ta} = 2.44$ \times $10$ ^{5} $ and (b) \textit{Ta} = 2.39$ \times $10$ ^{7} $. All results are normalized by the angular velocity flux of the nonvortical laminar state \textit{J}$ ^{\omega}_{lam} $, and only the data within the scope of \textit{r$ _{i} $}+$ \delta $ $ \le $ \textit{r} $ \le $ \textit{r$ _{o} $} are shown.}\label{Contributions to the torque}
\end{figure}

In TC flow, the angular velocity flux is calculated as \textit{J}$ ^{\omega} $ = \textit{r}$ ^{3} $($ \langle $\textit{u}$ _{r} $$ \omega$$\rangle$$_{A,t} $$ - $$ \nu $$ \partial $$ _{r} $$ \langle $$ \omega $$ \rangle $$ _{A,t} $), where the first term is the convective contribution and the second term is the diffusive (or viscous) contribution \citep{eckhardt2007torque}. The radial profiles of these two contributions for different rotating directions of the inner rough wall with $ \delta $ = 0.2\textit{d} at \textit{Ta} = 2.44$ \times $10$ ^{5} $ and \textit{Ta} = 2.39$ \times $10$ ^{7} $ are exemplified in figure \ref{Contributions to the torque}. It can be seen that the convective contribution to the torque is mainly in the central region and disappears at the boundaries, as expected. In contrast, the diffusive contribution dominates near the walls but drops to almost zero in the middle.

Furthermore, as shown in figure \ref{Contributions to the torque}(a), the rotating directions have no effect on the convective and diffusive contributions to \textit{J}$ ^{\omega} $ at low Taylor number \textit{Ta} = 2.44$ \times $10$ ^{5} $ which is in the laminar Taylor vortices regime.
On the other hand, as shown in figure \ref{Contributions to the torque}(b), the situation becomes different at a large Taylor number \textit{Ta} = 2.39$ \times $10$ ^{7} $ that corresponds to the turbulent Taylor vortices regime. It can be seen that the diffusive contribution is still unaffected by the rotating direction of inner rough wall except for the inner and outer boundaries, but the rotating direction has a significant effect on the convective contribution to the torque. When the inner wall rotates in the clockwise direction ($  \omega_{i} $ $ < $ 0), the convection term is larger than that of the counter-clockwise direction ($ \omega_{i} $ $ > $ 0). The results presented in figure \ref{Contributions to the torque} are consistent with those reported in figure \ref{NuvsTa} and show that the torque enhancement is dominantly due to the increased convective contribution.

\begin{figure}
	\centering
	{\includegraphics[width=0.7\textwidth]{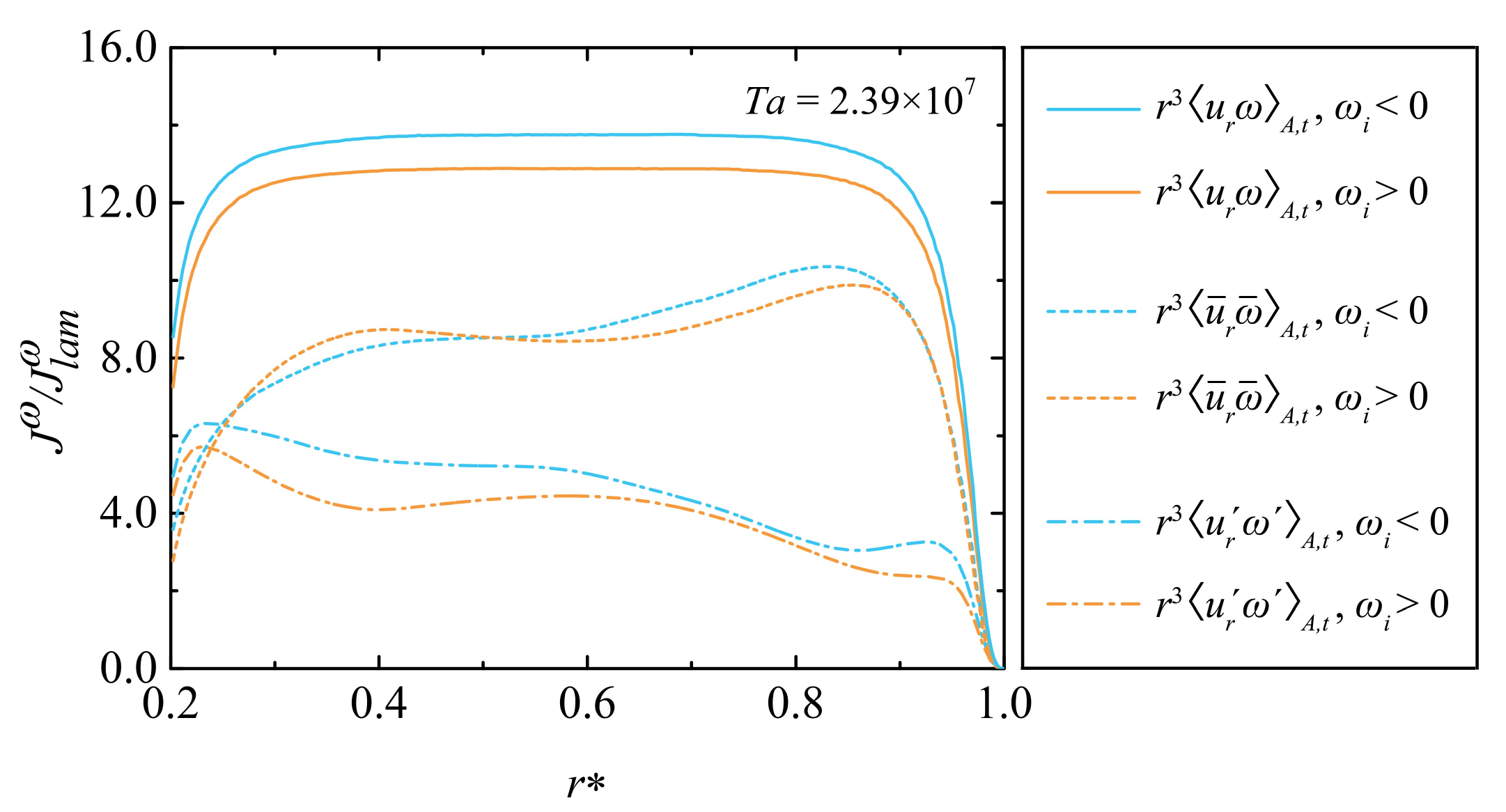}}
	\caption{The average and fluctuation contributions to the convective term of total angular velocity flux for different rotating directions with $ \delta $ = 0.2\textit{d} at \textit{Ta} = 2.39$ \times $10$ ^{7} $. All results are normalized by the angular velocity flux of the nonvortical laminar state \textit{J}$ ^{\omega}_{lam} $, and only the data within the scope of \textit{r$ _{i} $}+$ \delta $ $ \le $ \textit{r} $ \le $ \textit{r$ _{o} $} are shown.}\label{Reynoldsstress}
\end{figure}

To better explain why the rotating direction of the inner rough wall influences the convective term of torque at large Taylor number \textit{Ta} = 2.39$ \times $10$ ^{7} $ in figure 5(b). As shown in figure \ref{Reynoldsstress}, the convection contributions to the total torque are further decomposed into two components, \textit{r}$ ^{3} $$ \langle $$ \overline\textit{u} $$ _{r} $$ \overline\omega$$\rangle$$_{A,t} $ is the averages  (which have a structure, due to the presence of Taylor rolls), and \textit{r}$ ^{3} $$ \langle $$ \textit{u}^{\prime}_{r} $$ \omega^{\prime} $$\rangle$$_{A,t} $ is arising from the correlation of the fluctuations. Figure \ref{Reynoldsstress} shows that compared with the turbulent convective flux \textit{r}$ ^{3} $$ \langle $$ \textit{u}^{\prime}_{r} $$ \omega^{\prime} $$\rangle$$_{A,t} $ caused by the Reynolds stress, the mean convective flux \textit{r}$ ^{3} $$ \langle $$ \overline\textit{u} $$ _{r} $$ \overline\omega$$\rangle$$_{A,t} $ caused by the presence of mean Taylor vortices dominates to derive torques \citep{brauckmann2013direct}. And when the rough inner wall rotates in different directions, for the average term, the value of clockwise rotation $ \omega_{i} $ $ < $ 0 near the smooth outer wall is greater than the value of counter-clockwise rotation $ \omega_{i} $ $ > $ 0, while the opposite is true on the side near the inner wall. Which results in a small effect of the different rotation directions of the inner rough wall on the mean convective flux \textit{r}$ ^{3} $$ \langle $$ \overline\textit{u} $$ _{r} $$ \overline\omega$$\rangle$$_{A,t} $ contribution to the total flux \textit{J}$ ^{\omega} $. However, the situation becomes simple for the fluctuant term, \textit{r}$ ^{3} $$ \langle $$ \textit{u}^{\prime}_{r} $$ \omega^{\prime} $$\rangle$$_{A,t} $ with $  \omega_{i} $ $ < $ 0 is always greater than the value with $  \omega_{i} $ $ > $ 0 at the same radius, indicating that the turbulence caused by clockwise rotation ($  \omega_{i} $ $ < $ 0) is more intense than that of $  \omega_{i} $ $ > $ 0. Those facts lead to the observed torque enhancement and the different effects of rotating directions.

\subsection{The mechanism of torque enhancement}
\begin{figure}
	\centering
	{\includegraphics[width=1\textwidth]{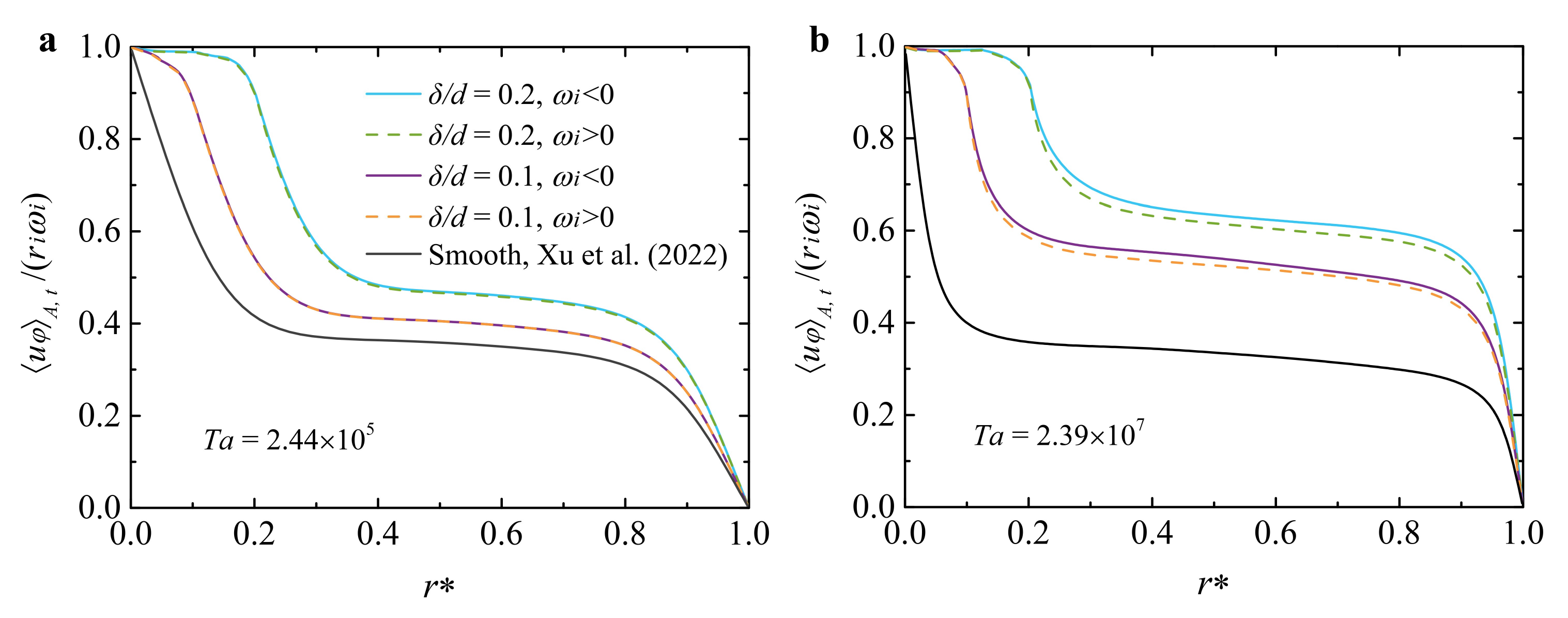}}
	\caption{Averaged azimuthal velocity profiles for different rotating directions of the inner rough wall with two strip heights at (a) \textit{Ta} = 2.44$ \times $10$ ^{5} $ and (b) \textit{Ta} = 2.39$ \times $10$ ^{7} $. The averaged azimuthal velocity profiles for smooth walls at these two \textit{Ta} from our previous study \citep{xu2022direct} are also included for comparison.}\label{Averaged azimuthal velocity}
\end{figure}

To understand the mechanism underlying the torque enhancement and the effect of rotating directions, it is useful to analyze the dependence of the azimuthal velocity profiles $ \textit{u}_{\varphi} $(\textit{r}) on the driving parameter \textit{Ta}. Therefore, $ \textit{u}_{\varphi} $(\textit{r}) for two representative Taylor numbers \textit{Ta} = 2.44$ \times $10$ ^{5} $ and \textit{Ta} = 2.39$ \times $10$ ^{7} $ are presented in figure \ref{Averaged azimuthal velocity}.
It can be seen that the azimuthal velocity profiles are influenced by the strip height, the higher the strip, the larger the azimuthal velocity at the same radius. For small Taylor number \textit{Ta} = 2.44$ \times $10$ ^{5} $, the azimuthal velocity profiles are almost unaffected by the rotating direction of the inner rough wall. At large Taylor number \textit{Ta} = 2.39$ \times $10$ ^{7} $, the rotating direction has an effect on the azimuthal velocity profiles, that is, the clockwise rotation $ \omega_{i} $ $ < $ 0 of the vertical asymmetric rough wall makes the azimuthal velocity larger at a given \textit{r}, compared with the case of counter-clockwise rotation $ \omega_{i} $ $ > $ 0.
This is because the shear rate of the azimuthal velocity at the rough wall is smaller than the one at the corresponding smooth case \citep{van2003smooth,zhu2017disentangling}, the azimuthal velocity should be biased towards the rough wall at the mid gap compared with the smooth case, leading to the stronger coupling between the rough wall and the bulk.
Furthermore, the difference in azimuthal velocity profiles with a higher strip height between counter-clockwise and clockwise rotation is larger.
These observations explain the torque enhancement and the effects of rotating direction to a certain extent.

\begin{figure}
	\centering
	{\includegraphics[width=1\textwidth]{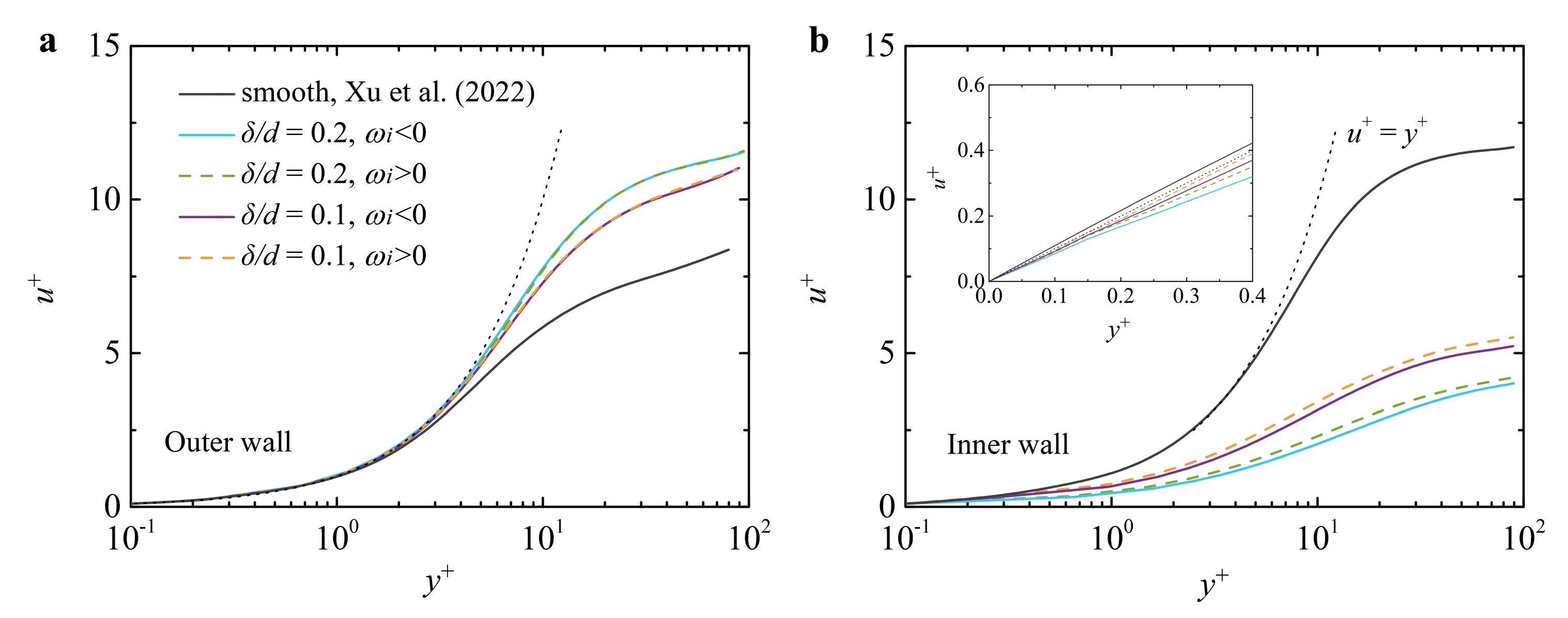}}
	\caption{The non-dimensionalized azimuthal velocity profiles for the outer stationary wall (a) and for different rotating directions of the inner rough wall with $ \delta $ = 0.1\textit{d} and $ \delta $ = 0.2\textit{d} (b) versus wall distance at \textit{Ta}=2.39$ \times $10$ ^{7} $. The inset in figure 8(b) is the enlargement near the inner wall. The non-dimensionalized azimuthal velocity profiles for smooth wall at the same \textit{Ta} are from our previous study \citep{xu2022direct}. For the outer smooth wall, the non-dimensionalized azimuthal velocity profile is \textit{u}$ ^{+} $ = $ \langle $$ u_{\varphi} $$ \rangle $$ _{A,t} $/\textit{u}$ _{\tau} $ and the wall distance is \textit{y}$ ^{+} $ = (\textit{r}$ _{o} $$ - $\textit{r})/$ \delta $$ _{v} $, where the friction velocity is \textit{u}$ _{\tau} $ = $\sqrt{\tau/2\pi\textit{l}\rho\textit{r}^{2}}$ = $\sqrt{Nu_{\omega}J^{\omega}_{lam}/\textit{r}^{2}}$, the boundary layer thickness $ \delta $$ _{v} $ is estimated by $ \delta $$ _{v} $ $ \approx $ $ d\sigma / (2Nu_{\omega})$, and $ \sigma $ is defined as $ \sigma = [(r_{i}+r_{o})/(2\sqrt{r_{o}r_{i}})]^{4} $ \citep{brauckmann2013direct,zhu2017disentangling}; for the inner rough wall, the non-dimensionalized azimuthal velocity profile \textit{u}$ ^{+} $ = ($ \overline\textit{u} $$ _{\varphi, r=r_{i}+\delta} $$ - $$ \langle $$ u_{\varphi} $$ \rangle $$ _{A,t} $)/\textit{u}$ _{\tau} $, the wall distance \textit{y}$ ^{+} $ = (\textit{r}$ - $\textit{r}$ _{i} $$ - $$ \delta $)/$ \delta $$ _{v} $. The dotted lines show the relationships \textit{u}$ ^{+} $ = \textit{y}$ ^{+} $.}\label{boundarylayer}
\end{figure}

It is well-known that the characteristics of velocity boundary layer (BL) reflect many features of wall turbulence \citep{grossmann2016high}. Therefore, the non-dimensionalized azimuthal velocity profiles \textit{u}$ ^{+} $ versus the wall distance \textit{y}$ ^{+} $ for the outer smooth wall and the inner rough wall in the case of \textit{Ta} = 2.39$ \times $10$ ^{7} $ are shown in figure \ref{boundarylayer}. Figure \ref{boundarylayer}(a) shows that there is a viscous sublayer (\textit{u}$ ^{+} $ = \textit{y}$ ^{+} $), 
which is well known for a smooth wall. 
The BL of the outer wall is influenced by the height of the strip attached to the inner wall, resulting in upward shifts of the log-law region, that is, the higher the strip, the larger the shift. However, the rotating direction of inner rough wall with the same strip height has no effect on the characteristics of velocity boundary layer near the outer stationary wall.
For the inner cylinder, it can be seen from figure \ref{boundarylayer}(b) that the BL is not only influenced by the strip height, but also affected by the rotating direction of the inner wall. Compared to the smooth inner wall, significant downward shifts are acquired for the rough cases, which are similar to the results of \cite{zhu2016direct}. Meanwhile, the downward trend is larger for the higher strip or for clockwise rotation ($ \omega_{i} $$ < $0) of the inner rough wall with a same strip height. It means that a higher strip and clockwise rotation ($  \omega_{i} $ $ < $ 0) of the inner rough wall can form a thinner BL at this \textit{Ta}. As a result, the torque enhancement becomes more obvious with the higher strip and clockwise rotation of the inner rough wall at large \textit{Ta}.

\begin{figure}
	\centering
	{\includegraphics[width=0.5\textwidth]{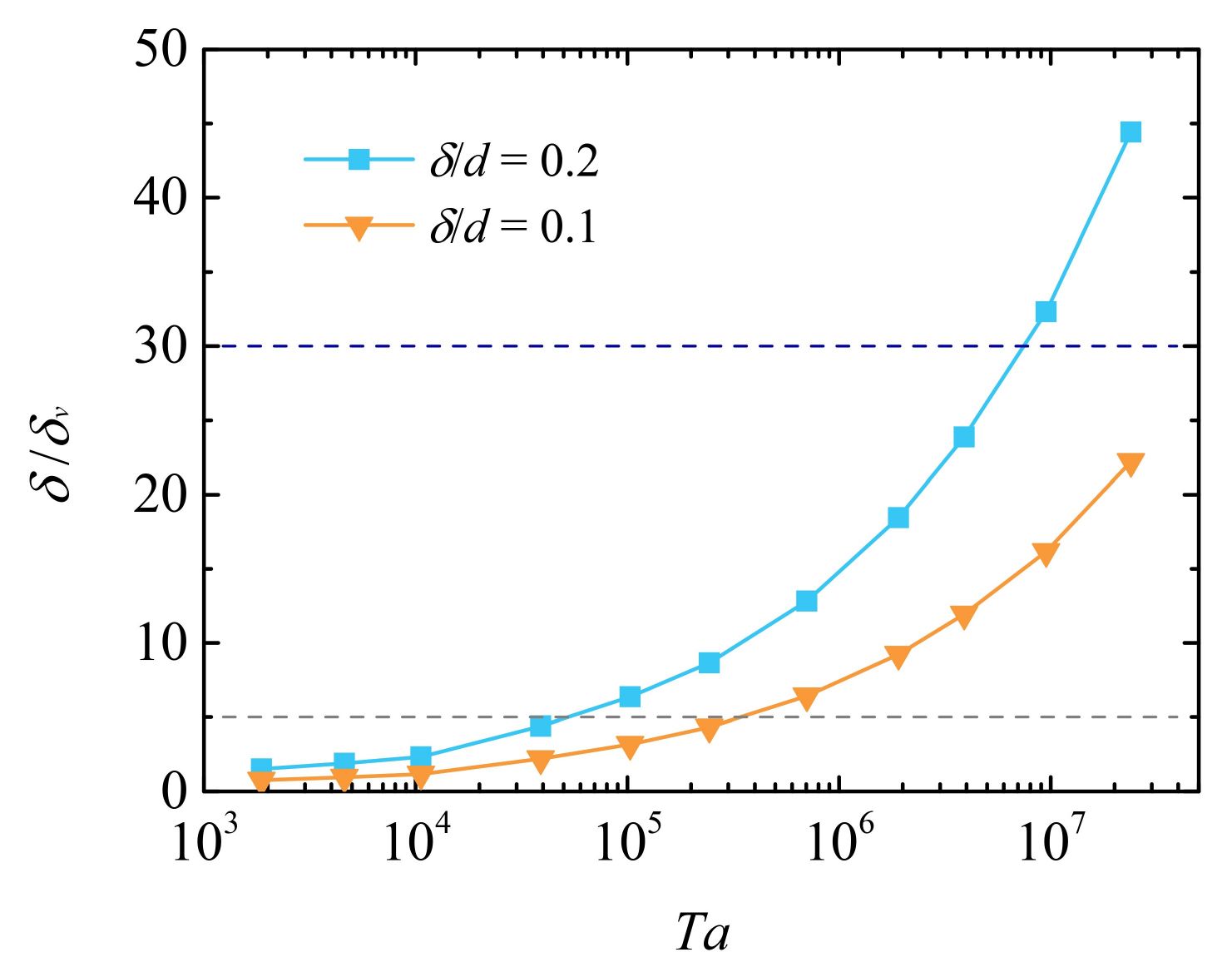}}
	\caption{The dimensionless value of the roughness height $ \delta $ in wall units. The gray and blue dashed lines denote the boundaries of the viscous sublayer ($ \delta $/$ \delta $$ _{v} $ $ < $ 5) and buffer layer (5 $ < $ $ \delta $/$ \delta $$ _{v} $ $ < $ 30).}\label{boundarylayerthickness}
\end{figure}

\begin{figure}
	\centering
	{\includegraphics[width=1\textwidth]{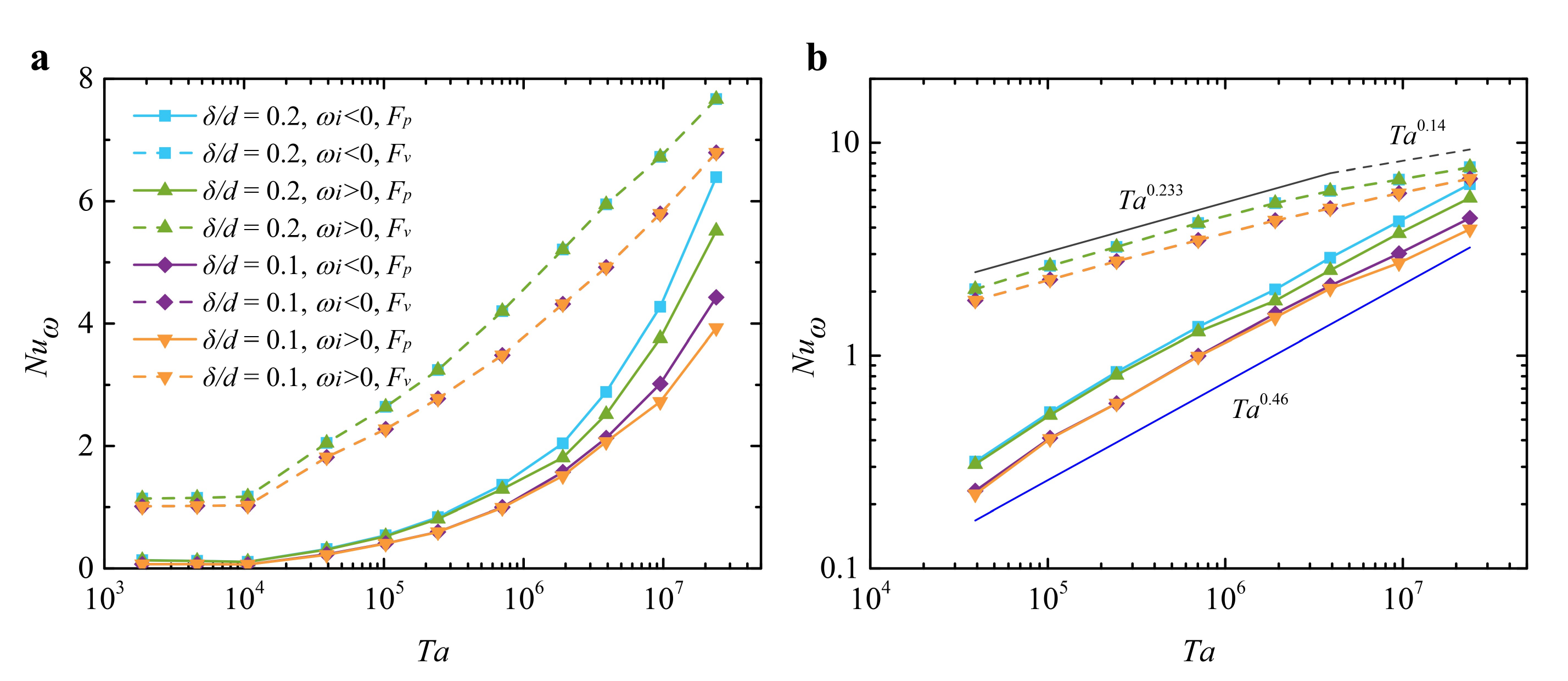}}
	\caption{(a) Contributions to the total torque originating from the pressure force \textit{Nu$ _{p} $} and the viscous force \textit{Nu$ _{\nu} $} with two strip heights $ \delta $ = 0.1\textit{d} and $ \delta $ = 0.2\textit{d} on the inner wall rotating in different directions. The solid lines represent the contribution of the pressure force \textit{F$ _{p} $} to the total torque, and the dash ones represent the contribution of the viscous force \textit{F$ _{\nu} $}. (b) The log-log scale of figure 10a.}\label{Pressure and viscous forces}
\end{figure}

We also plot the dimensionless value $ \delta $/$ \delta_{v} $ of the roughness height in wall units for different strip heights in figure \ref{boundarylayerthickness}, where the gray and blue dashed lines denote the boundaries of the viscous sublayer ($ \delta $/$ \delta_{v} $ $ < $ 5) and buffer layer (5 $ < $ $ \delta $/$ \delta_{v} $ $ < $ 30) \citep{pope2000turbulent}. Figure \ref{boundarylayerthickness} shows that if these dimensionless values are in the viscous sublayer, the flow is dominated by viscous effects and the rotation direction of the asymmetric rough wall has no change at all on the torque Nusselt number (which can be seen in figure 3b), which is similar with the common finding that the rough surface become active only the thermal boundary layer thickness is smaller than the characteristic height of roughness \citep{shen1996turbulent,stringano2006turbulent} in Rayleigh-B\'{e}nard (RB) flow. In TC flow with parallel grooves, same results reported by \cite{zhu2016direct} showed that the effect of grooves on the torque $ Nu_{\omega} $ can only be seen when the BL thickness becomes thinner than the groove height. On the other hand, the pressure drag is affected by the rotation direction of the asymmetric roughness in the buffer layers when $ \delta $/$ \delta_{v} $ $ \approx $ 16 for $ \delta $ = 0.1\textit{d} at \textit{Ta} = 9.52$ \times $10$ ^{6} $ and $ \delta $/$ \delta_{v} $ $ \approx $ 16 for $ \delta $ = 0.2\textit{d} at \textit{Ta} = 1.91$ \times $10$ ^{6} $ in present simulations (as shown in figure \ref{Pressure and viscous forces}b). Furthermore, it is worth noting that a 2D triangle with the top facing forward (into the wind) has a lower drag as compared to the same triangle with the top pointing in the downstream direction at sufficiently large Reynolds numbers (e.g. drag coefficient 1.6 versus 2.0 according to White, Fluid Mechanics, 2011, Table 7.2). Our results are consistent to \cite{white2011chapter}, albeit the Reynolds number is relatively lower and the triangles are attached to the wall.

In the present study, the TC system is driven by the rotation of the inner cylinder. To reveal the mechanism of torque enhancement more directly, it is necessary to study the torque at the inner wall.
To find the mechanism behind the increase of \textit{Nu}$ _{\omega} $ for the vertical strips on the inner wall, the pressure and viscous contributions at the rough wall are quantified. The part of pressure force is defined as \citep{zhu2017disentangling}
\begin{equation}\label{3}
	Nu_{p}=\int\frac{pr}{\tau_{pa}}dS,
\end{equation}
where \textit{p} is the pressure, \textit{r} is the radius, $ \tau_{pa} $ is the torque required to drive the system in the purely azimuthal and laminar flow. While the part of viscous force is defined as \citep{zhu2017disentangling}
\begin{equation}\label{3}
	Nu_{\nu}=\int\frac{\tau_{\nu}r}{\tau_{pa}}dS,
\end{equation}
where \textit{$ \tau_{\nu} $} is the viscous shear stress.

Figure \ref{Pressure and viscous forces}(a) shows the contributions to the total torque originating from \textit{Nu$ _{p} $} and \textit{Nu$ _{\nu} $} for asymmetric vertical rough wall rotating in different directions with two strip heights $ \delta $ = 0.1\textit{d} and $ \delta $ = 0.2\textit{d}, the log-log scale are also shown in figure \ref{Pressure and viscous forces}(b) at laminar and turbulent vortex regimes. As shown in figure \ref{Pressure and viscous forces}(a), at small Taylor numbers, the torque on the rough wall almost all comes from the viscous force. With increasing \textit{Ta}, the contributions of viscous and pressure forces to the torque both increase but the latter is significantly faster than the former. More importantly, \textit{Nu$ _{\nu} $} is independent of the rotating directions at a same strip height although the strip is asymmetric. Furthermore, the higher the strip, the larger the viscous force to the total torque at the same \textit{Ta}. By contrast, \textit{Nu$ _{p} $} in the clockwise rotation cases are larger than those for counter-clockwise rotation at the same strip height, which hasn't been seen in the previous study with symmetric rough walls \citep{zhu2017disentangling}. Those facts explain the results shown in figure \ref{NuvsTa}, that is, the torques of clockwise rotation becomes larger than those of counter-clockwise rotation for the same \textit{Ta} and indicate that the torque difference of different rotating directions is dominantly due to the different contribution of \textit{Nu$ _{p} $}. 

Furthermore, we also plot the log-log scale in Figure 10(b) from \textit{Ta} = 3.9$\times$10$^{7}$ to \textit{Ta} = 2.39$\times$10$^{7}$. This reveals that both the contribution of viscous forces to torque and the overall torque display segmented power-law relationships with respect to \textit{Ta}. In both scenarios, distinct power-law exponents emerge in the laminar and turbulent Taylor vortex regimes. Within the laminar regime, the power-law exponent is greater, signifying a more pronounced dependency of torque contribution on \textit{Ta}. Conversely, in the turbulent regime, the power-law exponent is smaller, indicating a comparatively weaker dependency. However, unlike the segmented power-law relationship between viscous forces and \textit{Ta}, the contribution of pressure to torque does not exhibit a segmented power-law relationship with \textit{Ta}. Instead, it is characterized by a smaller prefactor yet a larger power-law exponent. This suggests that pressure plays a more substantial role in torque generation, showing a stronger dependence on \textit{Ta}. This implication aligns with the dominance of pressure at high Taylor numbers, as reported by \cite{zhu2017disentangling}. In summary, these observations indicate the intricate relationships between viscous stress and pressure in different \textit{Ta} number regimes.

\section{Conclusions}

In the present study, extensive direct numerical simulations were conducted to explore the effect of inner rough walls on the transport properties of Taylor-Couette flow. The inner cylinder was roughened by attaching eighteen vertical asymmetric strips, with strip heights of $ \delta $ = 0.1\textit{d} and $ \delta $ = 0.2\textit{d}. Numerical results were obtained for \textit{Ta} ranging from 1.87$ \times $10$ ^{3} $ to 2.39$ \times $10$ ^{7} $ at a radius ratio of $ \eta $ = 0.714 and an aspect ratio of $ \varGamma $ = 2/3$ \pi $, using periodic boundary conditions in the azimuthal and axial directions.

The main conclusions that can be drawn include:
(i) The rotation direction of the vertical asymmetric rough wall has a negligible effect on the torque at low Taylor numbers. The influence gradually becomes more pronounced with increasing \textit{Ta}, and the drag enhancement effect of clockwise rotation of the inner cylinder is more significant than that of counterclockwise rotation;
(ii) The rotation direction of vertical asymmetric rough wall also has a negligible effect on the azimuthal velocity and the Reynolds stress at low Taylor number, they are however larger at high \textit{Ta} when the inner cylinder rotating in clockwise direction; 
(iii) For large \textit{Ta}, the velocity boundary layer (BL) in the case of clockwise rotation is thinner than that in the case of counterclockwise rotation, due to the observed stronger turbulence.
(iv) The torque on the vertical asymmetric rough wall is derived from the viscous force and the pressure force. The contribution of the viscous force to the torque with the same strip height is always equal at the same \textit{Ta}, irrespective of the rotating direction of the inner vertical asymmetric wall. On the other hand, the contribution of the pressure force to the torque for the same strip height is unaffected by the rotating direction of the inner wall at low Taylor numbers but is significantly affected at large \textit{Ta}. Moreover, the contribution of the pressure force in the case of clockwise rotation is larger than that in the case of counterclockwise rotation, resulting in the observed larger torque in clockwise rotation.

\section*{Acknowledgement}
This study is financially supported by National Natural Science Foundation of China (11988102, 21978295), and the New Cornerstone Science Foundation through the XPLORER prize.\\

\noindent
\textbf{Declaration of interests.} The authors report no conflict of interest.\\

\section*{Appendix: Resolution tests and numerical details}
To obtain reliable numerical results, the grid's spatial resolutions have to be sufficient. The requirements for spatial resolution is to have the grid length in each direction of the order of local Kolmogorov length. In present simulations, the hexahedral grid was uniform in the azimuthal and axial directions, and refined near the inner and outer cylindrical walls in the radial direction \citep{dong2007direct,ostilla2013optimal}. In TC flow, \textit{J}$ ^{\omega} $ and \textit{Nu}$ _{\omega} $ = \textit{J}$ ^{\omega} $/\textit{J}$ ^{\omega}_{lam} $ should not be a function of the radius as mentioned previously, but numerically it does show some dependence. Because of numerical error, \textit{J}$ ^{\omega} $ will deviate slightly along the \textit{r} from a fixed value. To quantify this difference, \cite{zhu2016direct} defined
\begin{equation}\label{1}
	\varDelta_{J} = \frac{max(J^{\omega}(r))-min(J^{\omega}(r))}{\langle J^{\omega}(r)\rangle_{r}},
\end{equation}
where the maximum and minimum values are determined over all \textit{r}, which is selected to be within the scope of \textit{r$ _{i} $}+$ \delta $ $ \le $ \textit{r} $ \le $ \textit{r$ _{o} $}. It is a very strict requirement for the meshes that $ \varDelta_{J} $ $ \le $ 0.01 \citep{ostilla2013optimal}. We make sure all the simulations meet this criterion, the details are listed in table 1.

\begin{figure}
	\centering
	{\includegraphics[width=1\textwidth]{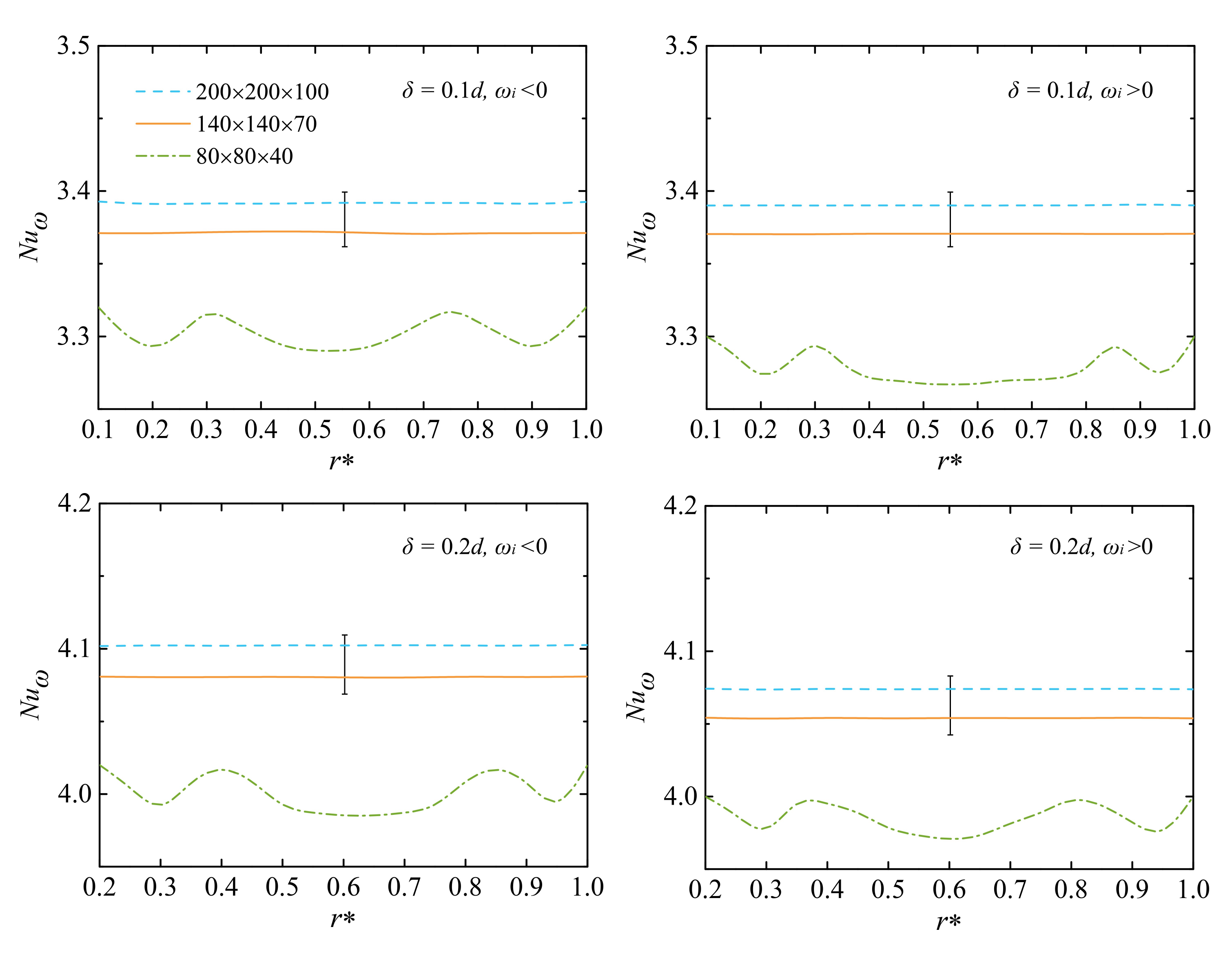}}
	\caption{Radial dependence of \textit{Nu}$ _{\omega} $ for different strip heights and different rotating directions of inner rough wall at \textit{Ta} = 2.44$ \times $10$ ^{5} $ with three different grid resolutions. An error bar indicating a 1\% error is provided for reference, the resolved cases lie within this error bar.}\label{Resolution tests}
\end{figure}

A resolution test of grid length has been exemplified in figure \ref{Resolution tests}, which presents four graphs of radial dependence of \textit{Nu}$ _{\omega} $ for different strip heights and different rotating directions of inner rough wall at \textit{Ta} = 2.44$ \times $10$ ^{5} $ with three different grid resolutions. An error bar indicating a 1\% error is provided for reference. It is shown that for the under-resolved cases ($N_{\varphi}$$ \times $$N_{r}$$ \times $$N_{z}$ = 80$ \times $80$ \times $40) the error of the \textit{Nu}$ _{\omega} $ along the radius is larger than 1\%. But the \textit{Nu}$ _{\omega} $ error is less than 1\% for the reasonably resolved cases ($N_{\varphi}$$ \times $$N_{r}$$ \times $$N_{z}$ = 140$ \times $140$ \times $70) and the extremely well-resolved cases ($N_{\varphi}$$ \times $$N_{r}$$ \times $$N_{z}$ = 200$ \times $200$ \times $100). 

\begin{table}
	\begin{center}
		\def~{\hphantom{0}}
		\begin{tabular}{lcccccc}
			\makebox[0.04\textwidth][l]{$ \delta/d $}&\makebox[0.15\textwidth][c]{$Ta$}&\makebox[0.08\textwidth][c]{$Re_{i}$}&\makebox[0.12\textwidth][c]{$Re_{\delta}$}&\makebox[0.2\textwidth][c]{$N_{\varphi}$$ \times $$N_{r}$$ \times $$N_{z}$}&\makebox[0.1\textwidth][c]{$Nu_{\omega}$}&\makebox[0.1\textwidth][c]{100$\varDelta_{J}$}\\[3pt]
			0.1   & 1.87$ \times $10$ ^{3} $ & $ - $35 &  $ - $3.25 & 90$ \times $90$ \times $45 & 1.08027 & 0.18\\
			0.1   & 1.87$ \times $10$ ^{3} $ & $ + $35 &  $ + $3.25 & 90$ \times $90$ \times $45 & 1.07885 & 0.22\\
			0.1   & 4.61$ \times $10$ ^{3} $ & $ - $55 &  $ - $5.19 & 100$ \times $100$ \times $50 & 1.09027 & 0.27\\
			0.1   & 4.61$ \times $10$ ^{3} $ & $ + $55 &  $ + $5.18 & 100$ \times $100$ \times $50 & 1.08859 & 0.25\\
			0.1   & 1.06$ \times $10$ ^{4} $ & $ - $83.5 &  $ - $7.94 & 110$ \times $110$ \times $55 & 1.09366 & 0.24\\
			0.1   & 1.06$ \times $10$ ^{4} $ & $ + $83.5 &  $ + $7.93 & 110$ \times $110$ \times $55 & 1.09224 & 0.31\\
			0.1   & 3.90$ \times $10$ ^{4} $ & $ - $160 &  $ - $14.49 & 120$ \times $120$ \times $60 & 2.04604 & 0.29\\
			0.1   & 3.90$ \times $10$ ^{4} $ & $ + $160 &  $ + $14.45 & 120$ \times $120$ \times $60 & 2.03780 & 0.25\\
			0.1   & 1.03$ \times $10$ ^{5} $ & $ - $260 &  $ - $22.63 & 130$ \times $130$ \times $65 & 2.68482 & 0.40\\
			0.1   & 1.03$ \times $10$ ^{5} $ & $ + $260 &  $ + $22.56 & 130$ \times $130$ \times $65 & 2.68105 & 0.36\\
			0.1   & 2.44$ \times $10$ ^{5} $ & $ - $400 &  $ - $33.51 & 140$ \times $140$ \times $70 & 3.37146 & 0.37\\
			0.1   & 2.44$ \times $10$ ^{5} $ & $ + $400 &  $ + $33.40 & 140$ \times $140$ \times $70 & 3.37098 & 0.41\\
			0.1   & 7.04$ \times $10$ ^{5} $ & $ - $680 &  $ - $55.53 & 160$ \times $160$ \times $80 & 4.48368 & 0.45\\
			0.1   & 7.04$ \times $10$ ^{5} $ & $ + $680 &  $ + $55.20 & 160$ \times $160$ \times $80 & 4.47725 & 0.42\\
			0.1   & 1.91$ \times $10$ ^{6} $ & $ - $1120 &  $ - $93.35 & 200$ \times $200$ \times $120 & 5.89129 & 0.57\\
			0.1   & 1.91$ \times $10$ ^{6} $ & $ + $1120 &  $ + $92.87 & 200$ \times $200$ \times $120 & 5.82758 & 0.54\\
			0.1   & 3.90$ \times $10$ ^{6} $ & $ - $1600 &  $ - $133.96 & 230$ \times $230$ \times $150 & 7.05058 & 0.66\\
			0.1   & 3.90$ \times $10$ ^{6} $ & $ + $1600 &  $ + $132.35 & 230$ \times $230$ \times $150 & 6.98768 & 0.71\\
			0.1   & 9.52$ \times $10$ ^{6} $ & $ - $2500 &  $ - $209.65 & 250$ \times $250$ \times $200 & 8.81297 & 0.79\\
			0.1   & 9.52$ \times $10$ ^{6} $ & $ + $2500 &  $ + $206.55 & 250$ \times $250$ \times $200 & 8.52065 & 0.72\\
			0.1   & 2.39$ \times $10$ ^{7} $ & $ - $3960 &  $ - $340.88 & 320$ \times $320$ \times $250 & 11.2194 & 0.80\\
			0.1   & 2.39$ \times $10$ ^{7} $ & $ + $3960 &  $ + $336.40 & 320$ \times $320$ \times $250 & 10.7176 & 0.85\\[3pt]
			0.2   & 1.87$ \times $10$ ^{3} $ & $ - $35 &  $ - $6.51 & 90$ \times $90$ \times $45 & 1.27208 & 0.21\\
			0.2   & 1.87$ \times $10$ ^{3} $ & $ + $35 &  $ + $6.50 & 90$ \times $90$ \times $45 & 1.26989 & 0.20\\
			0.2   & 4.61$ \times $10$ ^{3} $ & $ - $55 &  $ - $10.52 & 100$ \times $100$ \times $50 & 1.27398 & 0.26\\
			0.2   & 4.61$ \times $10$ ^{3} $ & $ + $55 &  $ + $10.51 & 100$ \times $100$ \times $50 & 1.27167 & 0.28\\
			0.2   & 1.06$ \times $10$ ^{4} $ & $ - $83.5 &  $ - $16.18 & 110$ \times $110$ \times $55 & 1.27882 & 0.24\\
			0.2   & 1.06$ \times $10$ ^{4} $ & $ + $83.5 &  $ + $16.14 & 110$ \times $110$ \times $55 & 1.27699 & 0.21\\
			0.2   & 3.90$ \times $10$ ^{4} $ & $ - $160 &  $ - $29.68 & 120$ \times $120$ \times $60 & 2.36599 & 0.23\\
			0.2   & 3.90$ \times $10$ ^{4} $ & $ + $160 &  $ + $29.57 & 120$ \times $120$ \times $60 & 2.35736 & 0.26\\
			0.2   & 1.03$ \times $10$ ^{5} $ & $ - $260 &  $ - $44.34 & 130$ \times $130$ \times $65 & 3.18239 & 0.28\\
			0.2   & 1.03$ \times $10$ ^{5} $ & $ + $260 &  $ + $44.09 & 130$ \times $130$ \times $65 & 3.16564 & 0.31\\
			0.2   & 2.44$ \times $10$ ^{5} $ & $ - $400 &  $ - $69.03 & 140$ \times $140$ \times $70 & 4.08044 & 0.30\\
			0.2   & 2.44$ \times $10$ ^{5} $ & $ + $400 &  $ + $68.40 & 140$ \times $140$ \times $70 & 4.05339 & 0.32\\
			0.2   & 7.04$ \times $10$ ^{5} $ & $ - $680 &  $ - $116.07 & 160$ \times $160$ \times $80 & 5.56394 & 0.36\\
			0.2   & 7.04$ \times $10$ ^{5} $ & $ + $680 &  $ + $114.57 & 160$ \times $160$ \times $80 & 5.49542 & 0.33\\
			0.2   & 1.91$ \times $10$ ^{6} $ & $ - $1120 &  $ - $187.66 & 200$ \times $200$ \times $120 & 7.25148 & 0.41\\
			0.2   & 1.91$ \times $10$ ^{6} $ & $ + $1120 &  $ + $185.92 & 200$ \times $200$ \times $120 & 7.01813 & 0.45\\
			0.2   & 3.90$ \times $10$ ^{6} $ & $ - $1600 &  $ - $278.16 & 230$ \times $230$ \times $150 & 8.83068 & 0.49\\
			0.2   & 3.90$ \times $10$ ^{6} $ & $ + $1600 &  $ + $271.54 & 230$ \times $230$ \times $150 & 8.46703 & 0.47\\
			0.2   & 9.52$ \times $10$ ^{6} $ & $ - $2500 &  $ - $449.22 & 250$ \times $250$ \times $200 & 10.9950 & 0.58\\
			0.2   & 9.52$ \times $10$ ^{6} $ & $ + $2500 &  $ + $439.74 & 250$ \times $250$ \times $200 & 10.4784 & 0.59\\
			0.2   & 2.39$ \times $10$ ^{7} $ & $ - $3960 &  $ - $725.56 & 320$ \times $320$ \times $250 & 14.0615 & 0.69\\
			0.2   & 2.39$ \times $10$ ^{7} $ & $ + $3960 &  $ + $712.28 & 320$ \times $320$ \times $250 & 13.1831 & 0.73\\
		\end{tabular}
		\caption{Values of the control parameters and the numerical results of the simulations. The columns display the strip height, the Taylor number, the inner Reynolds number, the Reynolds number for the roughness, the resolution employed, the dimensionless torque \textit{Nu}$ _{\omega} $ and the maximum deviation of angular velocity flux $ \varDelta $\textit{J}, respectively. All of the simulations are run in reduced geometry with $ \varGamma $=2$ \pi $/3 and a rotation symmetry of the order of six. The corresponding cases at the same \textit{Ta} without roughness (with smooth cylinders) can be found in our previous study \citep{xu2022direct} and in \cite{ostilla2013optimal}.}
		\label{table}
	\end{center}
\end{table}

\bibliographystyle{jfm}
\bibliography{jfm-instructions}

\end{document}